\documentclass[english,aps,prstper,reprint,showpacs,titlepage,longbibliography]{revtex4-2}   

\usepackage[T1]{fontenc}	
\usepackage[latin9]{inputenc}	
\usepackage{geometry}		
\usepackage{graphicx}
\usepackage[above,below]{placeins}	
\usepackage{times}
\usepackage{amsmath, amssymb, bm, mathtools, mathdots}

\usepackage{hyperref}  
\hypersetup{colorlinks=true,urlcolor=blue,citecolor=blue,linkcolor=blue}   
\usepackage{enumitem}          
\setlist{nosep}                 

\begin{document}

\begin{titlepage}

  \title{Comparing student performance on a multi-attempt asynchronous assessment to a single-attempt synchronous assessment in introductory level physics}

  \author{Emily Frederick (she/her)}
  \affiliation{Department of Physics, University of Central Florida, 4000 Central Florida Blvd, Orlando, FL 32816} 
  
  \author{Zhongzhou Chen (him/his)}
  \affiliation{Department of Physics, University of Central Florida, 4000 Central Florida Blvd, Orlando, FL 32816} 


  \begin{abstract}
    The current paper examines the possibility of replacing conventional synchronous single-attempt exam with more flexible and accessible multi-attempt asynchronous assessments in introductory-level physics by using large isomorphic problem banks. We compared student's performance on both numeric and conceptual problems administered on a multi-attempt, asynchronous quiz to their performance on isomorphic problems administered on a subsequent single-attempt, synchronous exam. We computed the phi coefficient and the McNemar's test statistic for the correlation matrix between paired problems on both assessments as a function of the number of attempts considered on the quiz. We found that for the conceptual problems, a multi-attempt quiz with five allowed attempts could potentially replace similar problems on a single-attempt exam, while there was a much weaker association for the numerical questions beyond two quiz attempts. 

    \clearpage
  \end{abstract}

  \maketitle
\end{titlepage}

\section{Introduction}
The majority of summative assessments in today's college Physics and STEM classes are conducted in the form of conventional single attempt synchronous exams, which require all students to take the exams at the same pre-determined time. However, this conventional assessment practice is becoming a significant accessibility hurdle for more and more students ~\cite{Zilles, Muldoon}, as the STEM higher-education student population becomes increasingly more diverse in every aspect ~\cite{USDOE}. Today's college students are coping with an increasingly wider range of disruptions, ranging from increasingly frequent climate-related natural disasters such as wildfire and hurricane, disease outbreaks such as COVID and flu ~\cite{Nguyen, Daniels, Zhang}, to managing work schedule conflicts and even childcare challenges. Conventional single attempt, synchronous assessment format is becoming an outstanding roadblock towards creating a more accessible and equitable learning environment for future students.

Several studies have shown the benefit of multi-attempt and asynchronous homework or assessment. For example, allowing students multi-attempts to practice on homework has shown an increase in learning ~\cite{Archer}, and asynchronous assessments increase accessibility and flexibility of assessments ~\cite{Hrastinski}. On asynchronous assessments, to prevent students from rote-memorization of problem answers or finding answers from external sources such as Chegg ~\cite{Persky, Mortati, Broemer}, most asynchronous assessments randomly select problems from problem banks of various sizes ~\cite{Silva, Chen}. While earlier multi-attempt assessments have been limited by the size of problem banks, recent advancements in applying large language models (LLMs) in education ~\cite{Kasneci} enables more efficient generation of problems. This advancement makes it possible to allow an even larger number of attempts on assessments. 

To evaluate this new assessment format, the current study tries to answer the question: \textit{To what extent can a multi-attempt, asynchronous, proctored assessment replaces a single attempt, synchronous proctored assessment?} More specifically, the current study tries to measure the following: for students who could answer a problem correctly within a given number of attempts on a multi-attempt, asynchronous, low stakes proctored assessment, what fraction could answer a similar problem on a subsequent single-attempt, synchronous, high stakes proctored exam? 

If both assessments were conducted under ideal conditions such that the students' performance reflects their true problem-solving ability, then their single attempt assessment performance should mostly reflect their last attempt performance on a multi-attempt assessment, regardless of the number of attempts allowed. This also assumes that students' problem solving ability improves after each failed attempt. In this study, we define performance as the score received on each problem. However, in practice, the association between students' performance on the two assessments could depend on at least the following three factors. 

First, allowing for too many attempts on a multi-attempt assessment could artificially increase the probability of obtaining a correct answer by chance, therefore over-estimating students' true ability. On the other hand, the single attempt assessment increases the chance of an accidental incorrect answer, leading to under-estimation of students' true ability. As a result, the highest correlation of performance between the two assessments would be detected when considering only correct answer made within a certain number of attempts on the multi-attempt assessment.

Second, some problem types such as multi-step numerical calculation problems could be more susceptible to the impact of unfavorable test conditions, due to more chances of making a mistake in one of the steps such as math calculation mistake. On the other hand, it may be easier to "guess" the answer of a multiple-choice problem on a multi-attempt assessment than a numerical problem.

Third, the association between the assessments will be weaker overall if the problems asked on the two assessments are less similar to each other.

Based on those impacting factors, the current study asks the following three specific research questions:
\begin{description}
    \item[RQ1] On which attempt is students' performance on the multi-attempt assessment most similar to their performance on a subsequent single-attempt assessment? 
    \item[RQ2] How does the above relation depend on the type of the problem being asked?
    \item[RQ3] How does the above relation depend on the level of similarity between the problems on both assessments?
\end{description}
 
We will answer these three questions by comparing students' performance on a multi-attempt assessment in the form of a quiz (referred to as quiz hereafter), to similar questions on a single attempt assessment in the form of a midterm exam (referred to exam hereafter). We will compare performance on both numerical calculation problems and conceptual problems. In the methods section, we will explain in detail how we measure the performance association between the two assessments. 

\section{Method}

\subsection{Instructional conditions}
We collected our data set from an undergraduate calculus based physics course taught by one of the authors at a large south-eastern public university in Fall 2023 semester, with 263 enrolled students. The course was taught in a blended mode with recorded online lectures and classroom activities alongside weekly discussions conducted by TAs. All course assessments were administered through the Canvas learning management system, including a total of five quizzes and three exams.  All quizzes were administered asynchronously and to be completed anytime during a one-week period. Students could choose to take the quiz either with webcam proctoring at a preferred location or during a discussion session with a TA present. The last two quizzes allowed a maximum of seven attempts using isomorphic problem banks.  

Each problem on the quiz was selected from a large isomorphic problem bank generated with the assistance of Large Language Model GPT-3 ~\cite{Playground}. We define isomorphic problems as problems that tests the same set of learning objectives, following a similar definition by Millar et al. ~\cite{Millar}. We generated the problem bank by first selecting a "seed" problem, and defining a set of viable isomorphic variations. We then prompted GPT-3.5 to generate new problems one at a time according to the seed problem and the pre-defined variations. Each generated problem would then be reviewed and edited by a human specialist to ensure validity. The process was repeated until the large problem bank was created. Solutions to each isomorphic problem was also created through the same process. The created problem banks were then uploaded to the learning management platform.

All three exams were synchronous, single attempt exams administered during regular class time. Students were either proctored by the instructor in the classroom or completed the exam at a different location at the same time being proctored through live webcam footage. 

\subsection{Study design and type of analysis}
This study chose to analyze student performance on problem banks implemented on quiz 4 and exam 3. Quiz 4 tests students' understanding of both conservation of linear momentum and conservation of mechanical energy, through one numerical problem and two conceptual problems. The numerical problem required students to input a numerical answer, and was selected from a bank that contains 44 problem variations. Each variation was isomorphic to the "ballistic pendulum" problem where a bullet hits a pendulum and then the pendulum (or the bullet and pendulum compound) is allowed to swing freely. Minor variations included replacing the bullets and pendulum with carts/boxes and while major variations require replacing the pendulum with an ideal spring. 

Both conceptual problems where multiple-answer questions with no calculation required, and were selected from a problem bank containing 21 variations. Each variation presents students a situation similar to the ballistic pendulum but asks which parts of the process conserved linear momentum and which parts conserved mechanical energy among the available options. Each problem has more than one correct answer. 

The exam consisted of ten problems, four of which were directly related to the quiz problems. Two of the four problems (one conceptual and one numerical) were selected directly from the quiz problem banks. We will refer to those problems as "original problems". These two original problems have two variations that were selected from each bank and randomly assigned to students. The other two problems (one conceptual, one numerical) were designed to test the same knowledge components using similar problem format, but not included in the open problem banks. We will refer to those as "transfer problems", as we would like to see if students are able to transfer their understanding to this new context.

We collected student performance data for all problems on quiz and the four related problems on exam 3. For the numerical problems, the outcome is recorded as binary (correct/incorrect). For the conceptual problems, the outcome include partial credits on each problem when students selected some but not all of the correct answers. From the performance data we extracted the following two types of information. 

First, students' performance on the quiz - this includes both the number of attempts needed to pass the quiz and the number of attempts needed to answer each type of problem correct on the quiz. The former is defined as the attempt number on which the student answered all the questions correctly (no student made new attempts after passing the quiz). The latter is defined as the attempt number on which the student first correctly answered a certain problem correctly, since in many cases not all problems were answered correctly on a given attempt. For the two conceptual problems, we record the attempt number on which both conceptual problems were answered correctly for the first time for each student.

Second, indicators of association between quiz and exam performance for a given number of quiz attempts. For each problem type, we could create a $2 \times 2$ correlation matrix for the outcome of the same type of problem on both the quiz and the exam, pairing the student performance into four distinct groupings as shown in an example table, Table \ref{tab1}. The number of students who got the same type of problem correct on both assessment are recorded in the AA group; the number of students who only got the problem on the quiz correct are recorded in the BA group, and so forth. We define "correct" on the quiz as correctly answering the problem within a given number of attempts $n$, for each problem types we can generate correlation matrices for $1 \leq n \leq 7$. 

\begin{table}[htbp]
  \caption{Example of the correlation matrix for the pairing of groups for comparing quiz and exam performance.\label{tab1}}
  \begin{ruledtabular}
    \begin{tabular}{ccc}
       & Quiz Correct & Quiz Incorrect \\ 
      \hline
      Exam Correct & AA & AB \\
      Exam Incorrect & BA & BB \\
    \end{tabular}
  \end{ruledtabular}
\end{table}

 For each matrix, we calculated the following two metrics:
\begin{enumerate}
  \item  Phi coefficient, defined as:
  \[
  \phi = \frac{AA\cdot BB - AB\cdot BA}{\sqrt{(AA+BA)(AB+BB)(AA+AB)(BA+BB)}}
  \] The coefficient measures the normalized difference between the diagonal terms and the off-diagonal terms. In the current context, it indicates the level of association between quiz performance and exam performance of the same problem type. Phi coefficient greater than 0.3 is often considered as moderate to strong correlation. 
  Its significance can be examined by chi-square test of association. A significant test result would indicate that the correlation between student responses on the two assessments are stronger than chance, when a certain maximum number of attempt is being considered on the quiz.
  \item McNemar test statistic ($\chi^2$): Defined as $\chi^2 = \frac{(AB-BA)^2}{AB+BA}$, this test statistic reflects the normalized magnitude of the difference between the two off-diagonal terms of the correlation matrix. In the current context, a significant test result would indicate that significantly more students obtained the correct answer to the question on one assessment compared to the other. A non-significant test result would indicate that it is equally likely for students to obtain correct answer of the problem on either assessment. 
\end{enumerate}
Statistical significance of both metrics in this study are reported after p-value adjustments \cite{Benjamini}.

\section{Results}
\subsection{Quiz and exam performance}
\begin{figure}
  \includegraphics[width=0.8\linewidth]{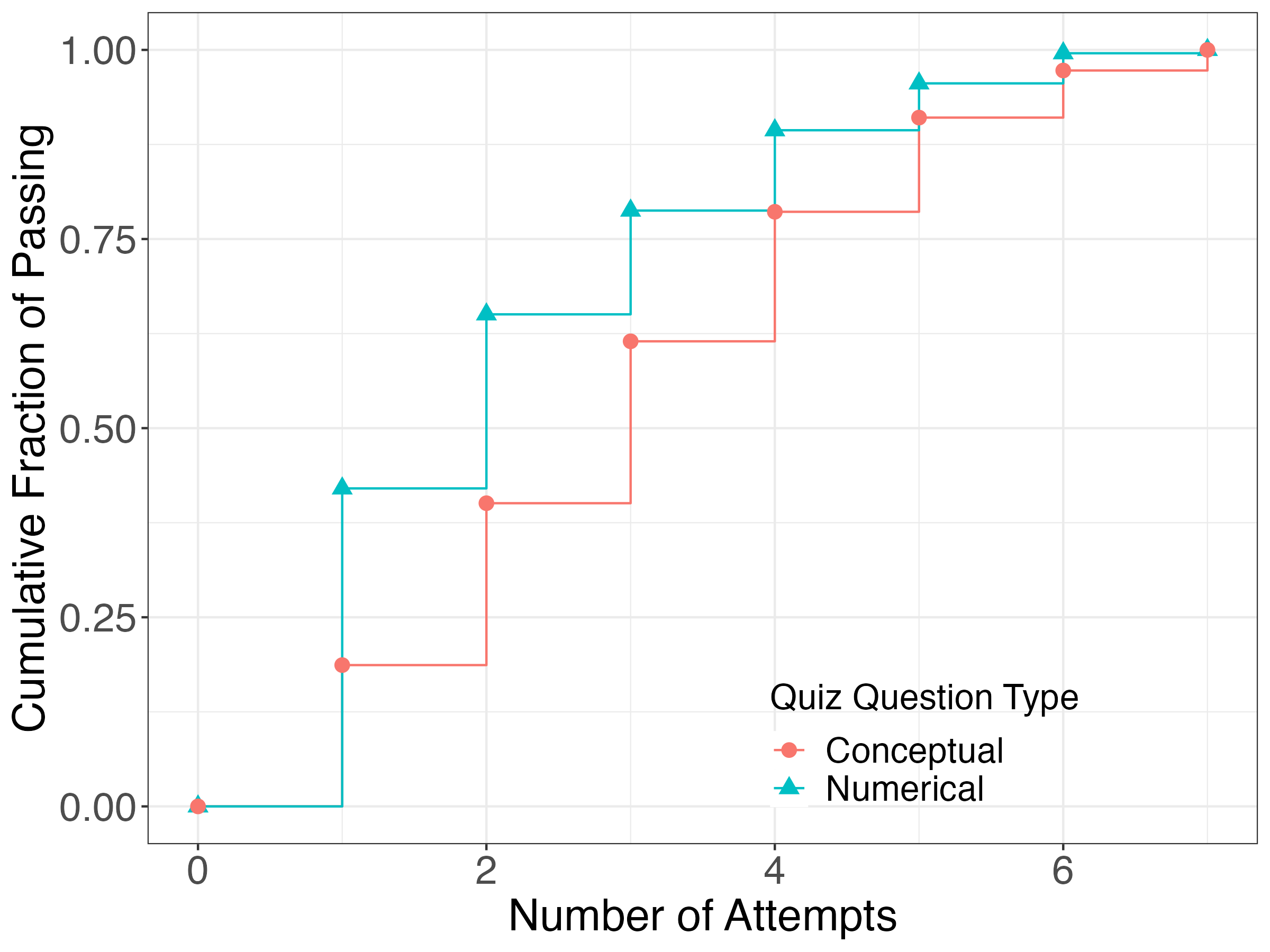}
  \caption{Cumulative fraction of a correct answer of the two types of questions (conceptual and numerical) on the multi-attempt quiz as a function of the number of attempts out of the students who passed the quiz.
  \label{figConNumQuiz}}
\end{figure}

A total of 262 students made at least one attempt on the quiz, and 128 (49\%) of those students passed the quiz within seven attempts (answered all problems correctly on one attempt.) No student attempted the quiz beyond their passing attempt. More than 90\%  of all students who passed the quiz did so within five attempts.  When examining the two question types separately, students are  likely to answer the numerical problem correctly one attempt earlier compared to answering both conceptual problem correctly (Fig. \ref{figConNumQuiz}).


A total of 263 students took the exam. Of which, 42\% answered the original numerical problem correctly, 30\% answered the transfer numerical problem correctly; 61\% answered the original conceptual problem correctly, where as only 34\% answered the transfer conceptual problem correctly. 

\subsection{Association between quiz performance and exam performance}
\begin{figure}
  \includegraphics[width=0.8\linewidth]{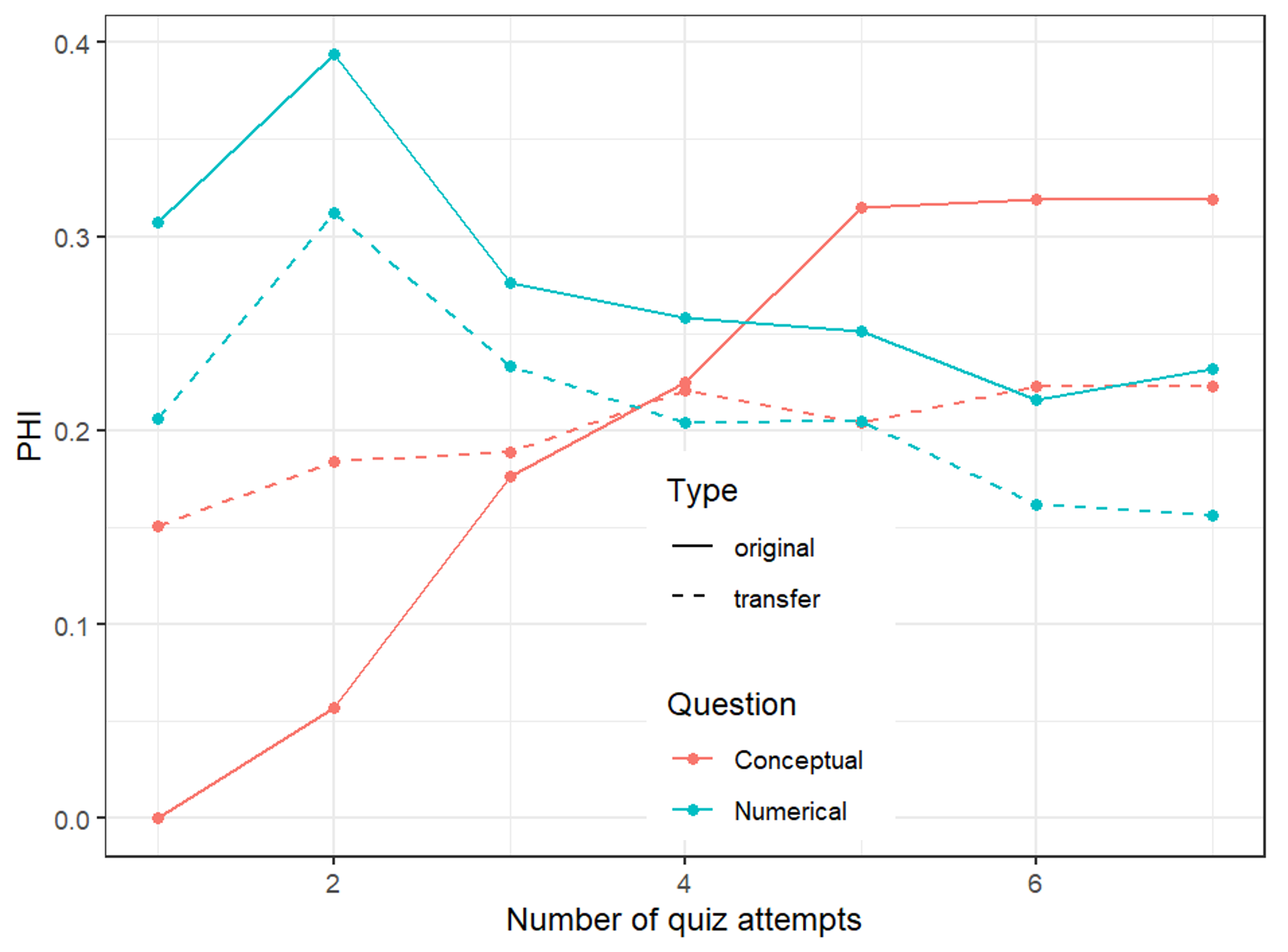}
  \caption{Phi coefficient between the exam performance and the quiz performance as a function of number of attempts on the quiz for the various types of questions combinations (Conceptual/numerical, Original/Transfer).  \label{figPhi}}
\end{figure}

\begin{figure}
  \includegraphics[width=0.8\linewidth]{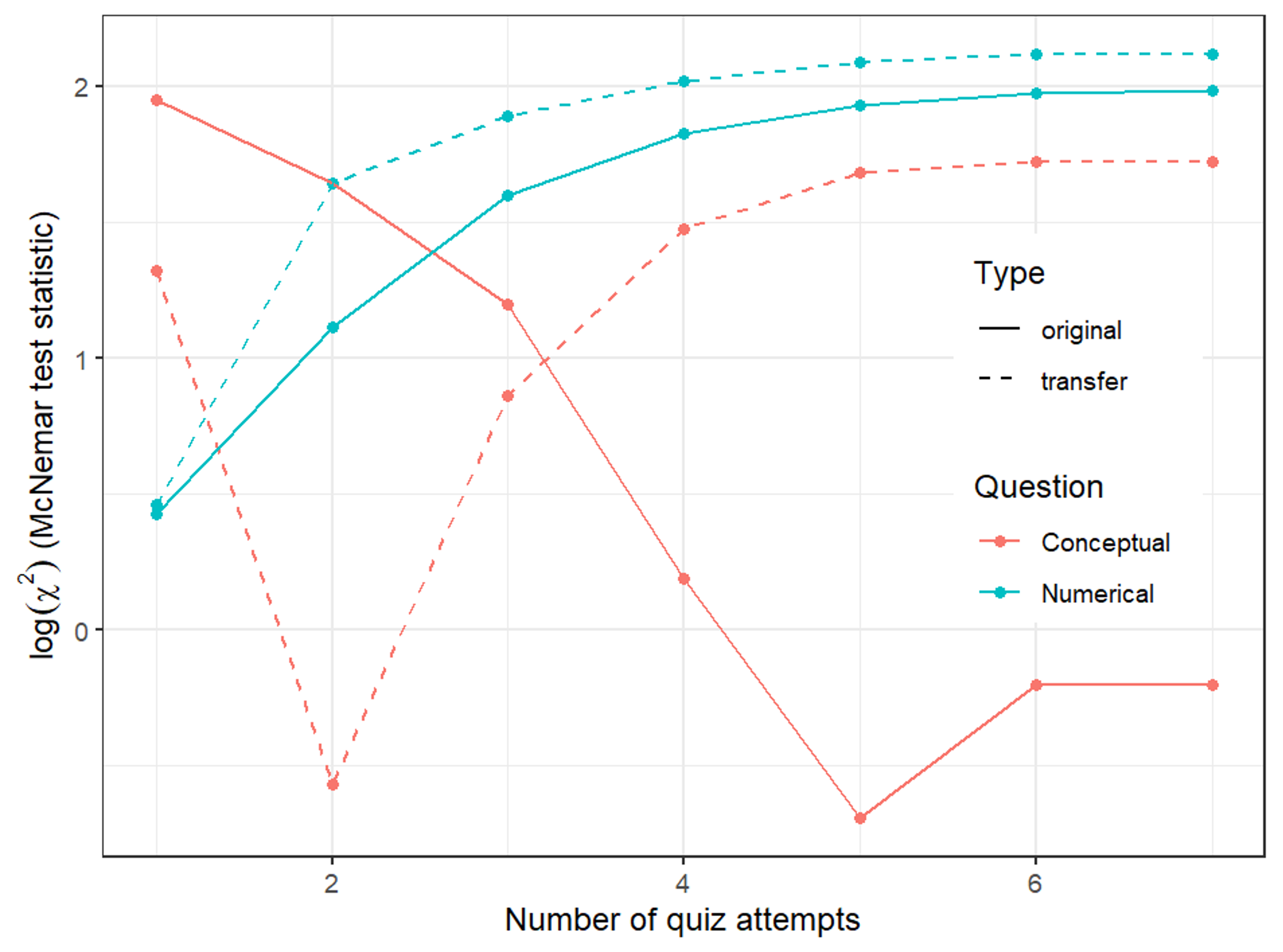}
  \caption{The log of test statistic ($\chi^2$) of McNemar's test plotted as a function of maximum numbers of attempts considered on the quiz.\label{figMcNemar}}
\end{figure}

We plot the Phi and $\chi^2$ indicies for the correlation matrix between the quiz and the exam in Fig. \ref{figPhi} and Fig. \ref{figMcNemar}, respectively. Both indices are plotted as a function of the maximum number of quiz attempts, meaning the maximum number of attempts under which students answered the type of problem correctly on the quiz. For example, the phi coefficient for numerical problem at attempt 3 is computed based on a correlation matrix for which a student is counted as "correct" on the quiz if they used three or less attempts to obtain the correct answer on the numerical problem on the quiz.

For the numerical questions (cyan), the relationship among the quiz problem to the exam original problem (solid) and to the exam transfer problem (dashed) exhibits similar trends. The phi coefficient is maximized when two quiz attempts are being considered, meaning that students' performance on the quiz is most similar to the exam, if we only consider their first two attempts. The association with the transfer numerical problem is weaker than that of the conceptual problems. All correlations are significant at $\alpha = 0.01$ level according to chi-square test, except for attempts six and seven for transfer problem, which are significant at $\alpha = 0.05$ level. However, the McNemar's test shows that the two assessments are unbalanced when more than one attempt is considered. Examining the correlation matrix shows that when more than one quiz attempt is allowed, it is easier to answer the numerical question correct on the quiz than on the exam. All McNemar test outcomes are significant at $\alpha = 0.01$ level beyond the one attempt. 

For the conceptual problems (orange), association with exam original problem is quite different from association with exam transfer problem. For the original conceptual question, the phi coefficient increases as more attempts are considered on the quiz up until five attempts. The McNemar's test statistic also reaches a minimum at five attempts and the test results are not significant beyond three attempts. Examining the correlation matrix revealed that the quiz was harder than the exam when less than five attempts were considered, and the two assessments reaches a balance at five or more attempts. For the transfer problem, the phi coefficient remains roughly the same for all attempts at around 0.2, with all correlation statistically significant at $\alpha = 0.01$ level. The McNemar's $\chi^2$ statistic has a minimum at attempt two, and the McNemar's test is significant at $\alpha = 0.01$ level for all other attempts. Examining the correlation matrix revealed that the quiz performance is better than the exam performance.

\section{Discussion and future directions}
We compared students' performance on a multi-attempt, asynchronous quiz to their performance on a single attempt, synchronous and found three distinct patterns of association.

First, for conceptual question that require students to answer two questions correctly on the quiz, students' quiz performance under five attempts is most similar to their exam performance on the original question, and the two assessments are of similar difficulty. Given that 90\% of the students who passed the quiz did so in fewer than five attempts, the evidences of this study suggest that it is feasible to replace a single-attempt, synchronous assessment with a multi-attempt, asynchronous assessment for this type of conceptual problem, as the impact of testing condition on either assessment is not dominant.

Second, when considering the transfer conceptual problem on the exam, the association is much weaker and the exam performance is significantly worse than the quiz when higher attempts are considered. A possible explanation is that students' learning from a failed quiz attempt is highly specific to the problem format, and generally lacks the ability to transfer to a new problem context. Some students may have also relied on memorization of problem answers. 

Third, for numerical questions, students' quiz performance within two attempts is most correlated to their exam performance. The association is weaker when three or more attempts are considered. In contrast, the quiz is significantly easier than the exam on both numerical problems when more than one attempt is considered. 

The lack of association between the numerical problems on more than two quiz attempts could suggest that less favorable test conditions such as limited time, challenging writing conditions, or inconvenient testing schedules which may have led to more student mistakes. Alternatively, the administration of the quiz could have introduced factors that made it artificially easy to obtain the correct answer to the problem. For example, some problems in the problem bank are easier than others for some students to solve. Numerical problems that are graded based only on the final answer seems to be more susceptible to non-ideal testing conditions. In future studies, these negative impacts could be mitigated by grading on students' written solution and assigning partial credit, and the extraneous factors of the quiz could be mitigated by requiring students to solve more than one numerical problem drawn from the problem bank. 

In conclusion, we found preliminary evidence that for conceptual problems, a five attempt asynchronous assessment that require students to correctly solve two isomorphic problems could replace a highly similar single attempt synchronous assessment, as the performance are most highly correlated. Students who failed on the fifth attempt should be required to study related content before being allowed to take additional attempts. For numerical problems, our results suggest that non-ideal assessment conditions might have significantly impacted students' performance. More work is needed to identify the impacting factors, reduce their impact for a more valid assessment, and evaluate whether one format produces more equitable assessment outcomes for different student demographics. Overall, we are hopeful that multi-attempt asynchronous assessments could potentially replace a significant fraction of synchronous single attempt assessments to provide more accessible and favorable assessments for students.

\acknowledgments
This study was supported by funding from NSF DUE-1845436 and University of Central Florida College of Science seed fund.


\begin{thebibliography}{0}%
\makeatletter
\providecommand \@ifxundefined [1]{%
 \@ifx{#1\undefined}
}%
\providecommand \@ifnum [1]{%
 \ifnum #1\expandafter \@firstoftwo
 \else \expandafter \@secondoftwo
 \fi
}%
\providecommand \@ifx [1]{%
 \ifx #1\expandafter \@firstoftwo
 \else \expandafter \@secondoftwo
 \fi
}%
\providecommand \natexlab [1]{#1}%
\providecommand \enquote  [1]{``#1''}%
\providecommand \bibnamefont  [1]{#1}%
\providecommand \bibfnamefont [1]{#1}%
\providecommand \citenamefont [1]{#1}%
\providecommand \href@noop [0]{\@secondoftwo}%
\providecommand \href [0]{\begingroup \@sanitize@url \@href}%
\providecommand \@href[1]{\@@startlink{#1}\@@href}%
\providecommand \@@href[1]{\endgroup#1\@@endlink}%
\providecommand \@sanitize@url [0]{\catcode `\\12\catcode `\$12\catcode `\&12\catcode `\#12\catcode `\^12\catcode `\_12\catcode `\%12\relax}%
\providecommand \@@startlink[1]{}%
\providecommand \@@endlink[0]{}%
\providecommand \url  [0]{\begingroup\@sanitize@url \@url }%
\providecommand \@url [1]{\endgroup\@href {#1}{\urlprefix }}%
\providecommand \urlprefix  [0]{URL }%
\providecommand \Eprint [0]{\href }%
\providecommand \doibase [0]{https://doi.org/}%
\providecommand \selectlanguage [0]{\@gobble}%
\providecommand \bibinfo  [0]{\@secondoftwo}%
\providecommand \bibfield  [0]{\@secondoftwo}%
\providecommand \translation [1]{[#1]}%
\providecommand \BibitemOpen [0]{}%
\providecommand \bibitemStop [0]{}%
\providecommand \bibitemNoStop [0]{.\EOS\space}%
\providecommand \EOS [0]{\spacefactor3000\relax}%
\providecommand \BibitemShut  [1]{\csname bibitem#1\endcsname}%
\let\auto@bib@innerbib\@empty
\end{thebibliography}%


\begin{thebibliography}{99}
  \bibitem{Zilles} C. Zilles, M. West, D. Mussulman, and T. Bretl, {\em Making Testing Less Trying: Lessons Learned from Operating a Computer-Based Testing Facility}, Proceedings - Frontiers in Education Conference, {\bf FIE 2018-Octob}, (2019).

  \bibitem{Muldoon} R. Muldoon, {\em Is It Time to Ditch the Traditional University Exam?}, Higher Education Research and Development {\bf 31}, 263 (2012).

  \bibitem{USDOE} U.S Department of Education and Office of Educational Technology, Reimagining the Role of Technology in Higher Education, 2017.

  \bibitem{Nguyen} T. Nguyen et al., {\em Insights into Students' Experiences and Perceptions of Remote Learning Methods: From the COVID-19 Pandemic to Best Practice for the Future}, Front Educ (Lausanne) 6, (2021).

  \bibitem{Daniels} L. M. Daniels, L. D. Goegan, and P. C. Parker, {\em The Impact of COVID-19 Triggered Changes to Instruction and Assessment on University Students' Self-Reported Motivation, Engagement and Perceptions}, Social Psychology of Education {\bf 24}, 299 (2021).

  \bibitem{Zhang} T. Zhang, M. Taub, and Z. Chen, {\em Measuring the Impact of COVID-19 Induced Campus Closure on Student Self-Regulated Learning in Physics Online Learning Modules}, in {\em LAK21: 11th International Learning Analytics and Knowledge Conference}, (ACM, New York, NY, USA, 2021), pp. 110-120. 

  \bibitem{Archer} K. Archer, {\em Do multiple homework attempts increase student learning? A quantitative study.}, The American Economist, 63 No. 2, (2018).

  \bibitem{Hrastinski} S. Hrastinski,  {\em Asynchronous and synchronous e-learning} Educause quarterly, {\bf 31} No.4, (2008).

  \bibitem{Kasneci} Kasneci, E., et al. {\em ChatGPT for good? On opportunities and challenges of large language models for education.} Learning and individual differences 103 (2023).

  \bibitem{Persky} A. M. Persky and K. A. Fuller, {\em Students' Collective Memory to Recall An Examination}, Am J Pharm Educ 8638 (2021)

  \bibitem{Mortati} J. Mortati and E. Carmel, {\em Can We Prevent a Technological Arms Race in University Student Cheating?}, Computer (Long Beach Calif) {\bf 54}, 90 (2021).

  \bibitem{Broemer} E. Broemer and G. Recktenwald, {\em Cheating and Chegg: A Retrospective}, ASEE Annual Conference and Exposition, Conference Proceedings, (2021).
  
  \bibitem{Silva} M. Silva, M. West, and C. Zilles. {\em Measuring the score advantage on asynchronous exams in an undergraduate CS course.} Proceedings of the 51st ACM Technical Symposium on Computer Science Education, (2020).

  \bibitem{Chen} B. Chen, M. West, and C. Zilles, {\em How much randomization is needed to deter collaborative cheating on asynchronous exams?} Proceedings of the 5th Annual ACM Conference on Learning at Scale, (2018). 
  
  \bibitem{Playground} \url{https://platform.openai.com/playground}. Retrieved
  05/04/2023.
  
  \bibitem{Millar} R. Millar and S. Manoharan, {\em Repeat Individualized Assessment Using Isomorphic Questions: A Novel Approach to Increase Peer Discussion and Learning}, International Journal of Educational Technology in Higher Education {\bf 18}, (2021).
  

  \bibitem{Benjamini} Y. Benjamini, and Y. Hochberg, {\em Controlling the false discovery rate: a practical and powerful approach to multiple testing}, Journal of the Royal Statistical Society Series B, {\bf 57}, (1995).

\end{thebibliography}

\end{document}